\documentclass[epjST]{svjour}

\usepackage{graphicx}

\usepackage{amsmath,amssymb}

\newcommand{\beq}{\begin{equation}}
\newcommand{\eeq}{\end{equation}}

\begin{document}
\title{Quantum Dynamics of Nuclear Slabs: Mean Field and Short-Range Correlations}
\author{Hossein Mahzoon\inst{1,2} \and Pawel Danielewicz\inst{1}\fnmsep\thanks{\email{danielewicz@nscl.msu.edu}} \and Arnau Rios\inst{3} }
\institute{Michigan State University, East Lansing, MI 48824, USA \and Truman State University, Kirksville, MO 63501, USA \and University of Surrey, Guildford, Surrey GU2 7XH, United Kingdom}
\abstract{
Computational difficulties aside, nonequilibrium Green’s functions appear ideally suited for investigating the dynamics of central nuclear reactions.  Many particles actively participate in those reactions.  At the two energy extremes for the collisions, the limiting cases of the Green’s function approach have been successful: the time-dependent Hartree-Fock theory at low energy and Boltzmann equation at high. The strategy for computational adaptation of the Green’s function to central reactions is discussed.  The strategy involves, in particular, incremental progression from one to three dimensions to develop and assess approximations,  discarding of far-away function elements, use of effective interactions and preparation of initial states for the reactions through adiabatic switching.  At this stage we concentrate on inclusion of correlations in one dimension, where relatively few approximations are needed, and we carry out reference calculations that can benchmark approximations needed for more dimensions.  We switch on short-range interactions generating the correlations adiabatically in the Kadanoff-Baym equations to arrive at correlated ground states for uniform matter.  As the energy of the correlated matter does not quite match the expectations for nuclear matter we add mean field to arrive at the match in energy.  From there on, we move to finite systems.  In switching on the correlations we observe emergence of extended tails in momentum distributions and evolution of single particle occupations away from 1 and~0.
} 
\maketitle
\section{Introduction}
\label{intro}
\begin{figure}
\centerline{
\includegraphics[width=.70\linewidth]{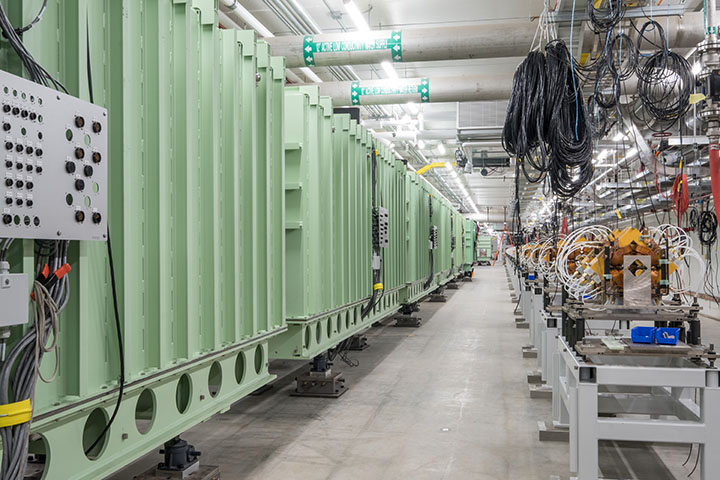}
}
\caption{Crymodules inside the tunnel of FRIB facility under construction at MSU.}
\label{fig:tunnel}       
\end{figure}

A major current factor expanding the scope of nuclear investigations is the construction of new accelerator facilities, in particular those accelerating secondary exotic beams, yielding new nuclides and reactions that could not be studied before.  A second high-intensity generation of the exotic-beam facilities is now under construction and these include the Facility for Rare Isotope Beams (FRIB) at the Michigan State University (MSU), at the moment the largest investment in the low-energy nuclear physics in North America, to be put into operation in year 2022, see Fig.~\ref{fig:tunnel}.  The expansion on the experimental side is paralleled by a growth in the theory for nuclear structure that, till now largely phenomenological, has increasingly become fundamentally based, even
down to QCD, as an effective theory.  On the side of nuclear reaction theory the progress has been much slower.  While fundamental advances were made for systems with few nucleons, particularly at low energy, the theory for central reactions of heavy nuclei, with many nucleons participating in the reaction dynamics, has remained phenomenological and largely semiclassical.  When many particles participate, it is natural for the theory to be statistical in nature.  For dynamics, the natural general way forward is the quantum transport that we will discuss here.

\section{Description of Central Nuclear Reactions}
\label{sec:1}
\begin{figure}
\centerline{
\includegraphics[width=.98\linewidth]{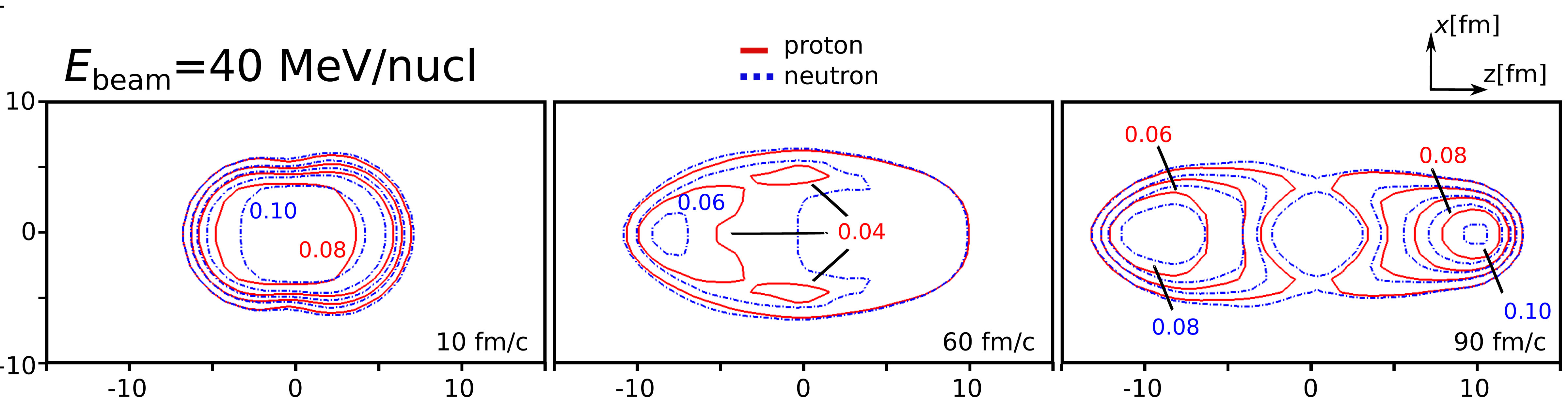}
}
\caption{Contour plots of neutron (dashed lines) and proton density (solid lines) in head-on $^{120}$Sn + $^{100}$Sn collision at $40\,\text{MeV/nucl}$, at different times \cite{stone_proton_2017}. The~horizontal axis is the collision axis.}
\label{fig:TDHF}       
\end{figure}

The number of methods developed for central nuclear reactions and employed in practice has been limited.  For low-energy reactions important is the time-dependent Hartree-Fock (TDHF) method \cite{Negele82}. Nowadays, simulations in variants of TDHF can be performed in full 3D and can involve nuclei as heavy as Uranium \cite{UMAR2015238,PhysRevLett.119.042501}. However, the validity of TDHF requires that role of correlations is suppressed in the dynamics~\cite{Tohyama87}. At low energies, one might argue that the role of correlations
is minimized by wavefunction antisymmetrization. Conversely, one would expect correlations to dominate at
higher energies, where role of the Pauli principle weakens.  With increase of collision energy the limitation of TDHF is exhibited in the fact that nuclei in TDHF simulations become excessively transparent to each other, see Fig.~\ref{fig:TDHF}, as compared to experiment.

Intermediate and high energy central nuclear reactions have been commonly described in terms of the Boltzmann-equation (BE) models \cite{bertsch88}.  In these models the evolution of the phase-space distribution functions $f({\pmb r},{\pmb p},t)$ of nucleons and other particles is followed.  BE models have been fairly successful in describing many aspects of higher-energy reactions \cite{danielewicz_production_1991,danielewicz_hadronic_2002}.  However, the use of BE in reactions has been criticized on theoretical grounds.  BE relies on the quasiparticle picture and simple estimates \cite{danielewicz84} indicate that particle scattering rates are comparable to particle energies, which undermines that picture.  In this context, it is theoretically difficult~\cite{dickhoff98} to separate collisional effects, described with cross-sections and entering the collisional integrals in BE from mean-field effects entering the quasiparticle energies.

\section{Kadanoff-Baym Equations}
\label{sec:KBeqs}

The TDHF and BE approaches to central nuclear reactions are fundamentally tied with the single-particle density matrix
\beq
\rho({\pmb r}_1 \, {\pmb r}_{1}' \, t)
= \langle \Phi | \psi^\dagger ( {\pmb r}_1' \, t) \, \psi ( {\pmb r}_1 \, t) | \Phi \rangle \, .
\label{eq:GF<}
\eeq
The density matrix yields all single-particle observables.  The Wigner function, that may be viewed as the quantal version of the classical phase-space distribution, results from evaluating a Fourier transformation of the equal-time Green's function, i.e.\ density matrix, in relative coordinates:
\beq
f({\pmb p} \, {\pmb r} \, t) = \int \text{d} ({\pmb r}_1 - {\pmb r}_1') \, \text{e}^{-i {\pmb p} ({\pmb r}_1 - {\pmb r}_1')} \, \rho({\pmb r}_1 \, {\pmb r}_{1}' \, t) \, .
\eeq
Here, the spatial argument of the Wigner function is ${\pmb r} = ({\pmb r}_1 + {\pmb r}_{1}')/2$.  It can be seen that the doubled spatial argument in the density matrix accounts for position and momentum in the classical limit.  By extension, one can expect that the doubled time argument in the Green's function corresponds to time and energy in the classical limit, i.e.\ the density in momentum {\em and} energy at a given position and time should be:
\beq
-iG^< ({\pmb p} \, \epsilon \, {\pmb r} \, t) =
\int \text{d} ({\pmb r}_1 - {\pmb r}_1') \, \text{d} (t_1 - t_1') \,
\text{e}^{i[\epsilon (t_1 - t_1')- {\pmb p} ({\pmb r}_1 - {\pmb r}_1')]} \,
(-i) G^<({\pmb r}_1 \, t_1 \, {\pmb r}_{1}' \, t_1') \, .
\label{eq:SpectralF}
\eeq
Indeed, for the static case of a Hartree-Fock state, where the Green's function is generally a superposition of products of occupied orbitals $\phi_\alpha$,
\beq
-iG^<({\pmb r}_1 \, t_1 \, {\pmb r}_{1}' \, t_1') = \sum_\alpha \phi_\alpha ({\pmb r}_1 \, t_1) \,
\phi_\alpha^* ({\pmb r}_1' \, t_1') \, ;
\eeq
Eq.~\eqref{eq:SpectralF} yields
\beq
-iG^< ({\pmb p} \, \epsilon \, {\pmb r} \, t) =  \sum_\alpha f_\alpha({\pmb p} \, {\pmb r}) \, \delta(\epsilon - \epsilon_\alpha) \, .
\eeq
Here, $\epsilon_\alpha$ are single-particle energies of the orbitals.  In nuclear physics, the spectral function represented by Eq.~\eqref{eq:SpectralF} is explored for ground-state nuclei in inelastic electron scattering.

Outside of the mean-field approximation, an intrinsically consistent dynamics for the Green's function \eqref{eq:GF<}, in nonequilibrium or even finite temperature situation, can only be arrived when considering different orderings of single-particle operators in an expectation value, encompassed in the Green's function
\beq
i \, G(1,1') = \langle \Phi | T \left\lbrace \psi (1) \, \psi^\dagger (1')   \right\rbrace    | \Phi \rangle \, .
\label{eq:GF}
\eeq
Here, the time arguments of the operators are assigned to either side of a time contour that runs first forward and then backward in time, representing the evolution of the ket in the expectation value, on one hand, and of the bra, on the other \cite{danielewicz84,kadanoff}.  The superoperator $T$ orders operators according to their order on the time contour.  Depending on the assignment of time arguments to the contour branches, different actual ordering of the operators can result, with the superscript '$<$' in the Green's function referring to the order in \eqref{eq:GF<} and '$>$' to the opposite order.

A perturbative expansion of the evolution operators in the Green's function \eqref{eq:GF}, followed by resummations, yields a Dyson equation familiar from other contexts
\beq
G = G_0 + G_0 \, \Sigma \, G \, .
\eeq
Here, the time integrations run over the contour forward and then backward in time, reaching above the maximal time of operators in any considered expectation value.  The self energy $\Sigma$ can be expressed in terms of source operators $j$,
\beq
\left( i \, \frac{\partial}{\partial t_1} + \frac{{\pmb \nabla}_1^2}{2 m} \right)  \psi(1) = j(1) \, ,
\eeq
with
\beq
i \, \Sigma(1,1') = \langle \Phi | T \left\lbrace j (1) \, j^\dagger (1')   \right\rbrace    | \Phi \rangle_\text{irr} \, .
\eeq

The integral Dyson equation may be converted into an integro-differential equation through application of the differential inverse operator $G_0^{-1}$ to both sides of the equation.  For a specific order of operators in $G$ \eqref{eq:GF}, one arrives at the set of Kadanoff-Baym equations \cite{kadanoff}:
\beq
\left( i \, \frac{\partial}{\partial t_1} + \frac{{\pmb \nabla}_1^2}{2 m} \right) \, G^\lessgtr (1, 1') =  \int \text{d} 1'' \, \Sigma^+ (1, 1'') \, G^\lessgtr (1'', 1')  +
\int \text{d} 1'' \, \Sigma^\lessgtr (1, 1'') \, G^- (1'', 1') \, .
\eeq
Implicit, in the expansion and resummation leading to the Dyson equation and, thus, to the Kadanoff-Baym (KB) equations, is the presumption that the impact of any multi-particle correlations dies off with time.

In different limits, for a variety of cases of $\Sigma$, the K-B equations yield a variety of useful known results and moreover they interpolate and extrapolate beyond those limits and associated results.  Thus, when the mean field contribution dominates the self-energy, $\Sigma_\text{mf} >> \Sigma^\lessgtr$, in a highly degenerate system, the TDHF description for the system follows. When the scales of variation for the Green's functions and self-energies are significantly larger for the average than relative function arguments, then the quasiparticle approximation follows, with evolution governed by the Boltzmann equation,
\beq
-i \, G^< (1,1') \approx \int \text{d} {\pmb p} \, f({\pmb p}, 1) \, \text{e}^{i \, {\pmb p} ({\pmb x}_1 - {\pmb x}_{1'}) - i \, \epsilon_{\pmb p} (t_{1}-t_{1'}) } \, .
\eeq

Given the two limiting approaches contained in the KB equations, already employed in the practice of central nuclear reactions, it is natural to ponder a direct application of the nonequilibrium Green's functions (NGF) to the reactions, covering the variety of circumstances that are possibly or clearly out of reach of those limiting approaches.  The problem with a direct solution of the KB equations, though, is the 8-dimensional nature of the functions that can easily overwhelm the current and near-future computational power.  By contrast, the TDHF equations involve 4 dimensions, or 5 if one counts the large number of orbitals, and these already tax the current power.

\section{Towards Reaction Simulations}

\subsection{General Challenges}
 General issues that must be considered when coping with nonuniform systems, described in terms of NGF, include the space-time matrix form of the dynamics, with abundance of matrix elements that nominally need to be considered.  Thus, for a coarse discretization
 involving sensible 50 values in any space-time direction, the number of matrix elements that would need to be considered is $50^8 = 4 \times 10^{13}$ for every function, of the order of $100 \, \text{TB}$ of data!   The coarse discretization in itself implies that effective rather than microscopic interactions need to be employed.  When dealing with time development, consistent approximations must be employed for initialized states and for dynamics, to avoid spurious evolution due to inconsistencies.

In order to progress on those challenges, we start with dynamics in one dimension.  In one dimension (1D), one can handle the number of elements for such a discretization as above and one can test approximations needed when moving to higher dimensionality.  Still, one needs to deal with effective interactions and initialization of dynamics.  In \cite{rios_towards_2011} we studied mean-field dynamics in 1D and demonstrated that only matrix elements of functions close to diagonal matter for evolution forward in time.  The far off-diagonal elements record phase relation between portions of the system widely separated in space.  When the system expands into vacuum those phase relations become irrelevant and can be discarded in evolution.  Here we go beyond mean field and concentrate on two issues: preparation of correlated initial state and effective interactions.

\subsection{Initialization of Correlated States}

In Sec.~\ref{sec:KBeqs} the KB equations, which govern the time evolution of a many-body system, were introduced. Solving the latter equations yields one-body information about the system and also many-body provided the effects of correlations are short-lived.  In studying collisions, nuclear systems need to be initialized consistently with the KB equations used to follow the collisions.  Thus, if an uncorrelated initial state is naively followed with KB equations incorporating correlations, then that system may spew particles and even explode and/or violently oscillate ahead of any collision.  In the past work~\cite{rios_towards_2011}, we employed adiabatic switching to arrive at mean-field initial states consistent with a mean-field version of the KB equations.  Examples, where adiabatic switching was used for constructing correlated many-body states, include Refs.~\cite{tohyama_stationary_1994,pfitzner_vibrations_1994}.  In \cite{danielewicz84b}, imaginary-time evolution was employed, requiring development of a separate computational infrastructure, but equivalent to the adiabatic switching under proper circumstances.

\subsection{Effective Interactions}

\begin{figure}
\centerline{
\includegraphics[width=.55\linewidth]{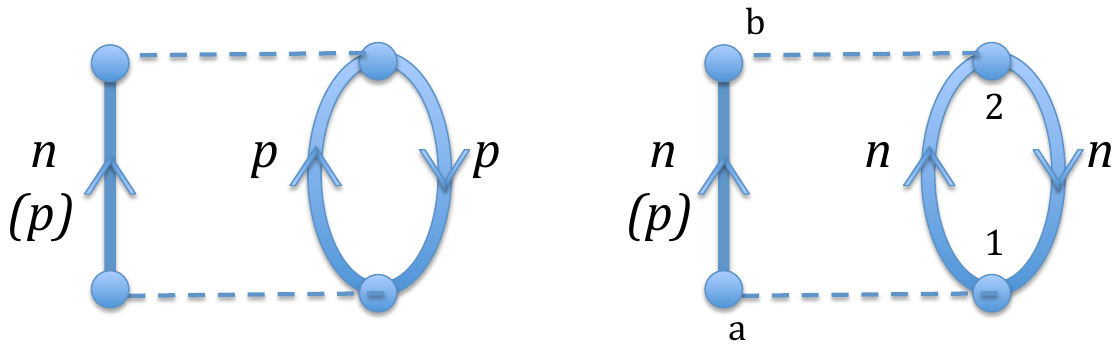}
}
\caption{Direct Born diagram contributions to the self energy for neutrons (protons).}
\label{fig:borniso}       
\end{figure}

The effective interactions, employed for coarse discretizations in space and time, need to be eventually derived through renormalization from effective interactions.  In 1D we do not have any microscopic interactions to start with, so we progress with effective interactions pragmatically, requiring that results obtained in 1D can be sensibly interpreted as pertaining to a three-dimensional (3D) nuclear system that is uniform in two perpendicular directions.
For the mean-field part of the self-energy, we adopt a local 3D Skyrme-type interaction such as in Ref.~\cite{rios_towards_2011}.  For the self-energies $\Sigma{\gtrless}$, we employ the so-called self-consistent Born diagram illustrated in Fig.~\ref{fig:borniso}, yielding
\begin{equation}
\Sigma^{\gtrless} (p \, t;p' \, t') =  \int \frac{\text{d} p_1}{2\pi}\frac{\text{d} p_2}{2\pi} \; V(p-p_1) \, V(p'-p_2)\; G^{\gtrless}(p_1 \, t;p_2 \, t')\, \Pi^{\gtrless}(p-p_1 \, t; p'-p_2 \, t') \, ,
\end{equation}
where
\begin{equation}
\Pi^{\gtrless}(p \, t;p' \, t') =  \int \frac{\text{d}p_1}{2\pi}\frac{\text{d}p_2}{2\pi}G^{\gtrless}(p_1 \, t;p_2 \, t') \, G^{\gtrless}(p_2-p' \, t';p_1-p \, t) \, .
\end{equation}

In the semiclassical low-density limit, the self-energies $\Sigma{\gtrless}$ represent phase-space feeding and depletion rates where cross sections are described in the Born approximation in terms of the employed residual interaction.  To describe semi-quantitatively 3D rates within the 1D calculation, we employ a residual interaction modulated by a Gaussian:
\begin{equation}
\label{eq:Vres}
\begin{aligned}
V(p) =& V_0 \, \sqrt{\pi} \, \eta^2 \, p^2 \, \text{e}^{-{\eta^2 \, p^2}/{4}} \, , \\
V(x) =& V_0 \, \left ( 1-2\frac{x^2}{\eta^2} \right ) \, \text{e}^{-{x^2}/{\eta^2}} \, .
\end{aligned}
\end{equation}
The constants $V_0$ and $\eta$ are adjusted to yield reasonable cross sections and quasiparticle content of single-particle states in the ground state.

\subsection{Adiabatic Transformation of Interactions}

The finalizing of the details of effective interactions is entangled in practice with arrival at static solutions of dynamics.  To ensure that the initial states are consistently determined with intentions for the dynamics, we arrive at states that are approximately static through adiabatic transformation of interactions within the dynamics.  We carry that out for uniform matter and for finite systems.  In considering the uniform matter, we adjust the characteristics of the mean-field self-energy $\Sigma_\text{mf}$, so that energy per nucleon, as a function of density, properly represents uniform 3D matter.

Central role in adiabatic switching is played by a switching function $F(t)$ that changes between 1 and 0 around some switching time $t_s$.  We start from an uncorrelated state and the residual interaction $V(x)$ is replaced by time-dependent interaction in the dynamics,
\beq
V_t(x,t) = [1-F(t)] \, V(x) \, ,
\eeq
and the mean-field self energy $\Sigma_\text{mf}$ is replaced by
\begin{equation}
U_t (x \, t) = F(t)\, U_0(x) +[1-F(t)] \, \Sigma_\text{HF} (x \, t) \, .
\end{equation}
For a finite system, a harmonic oscillator (HO) potential $U_0(x) = \frac 1 2 m \Omega^2 x^2$ is used to confine the original state.  For a uniform system local energy or potential plays a passive role in the dynamics, just changing the net energy by some value, so in practice $\Sigma_\text{mf}$ may be left to be specified after conclusion of evolution.

\begin{figure}
  \centerline{\includegraphics[width=.7\textwidth]{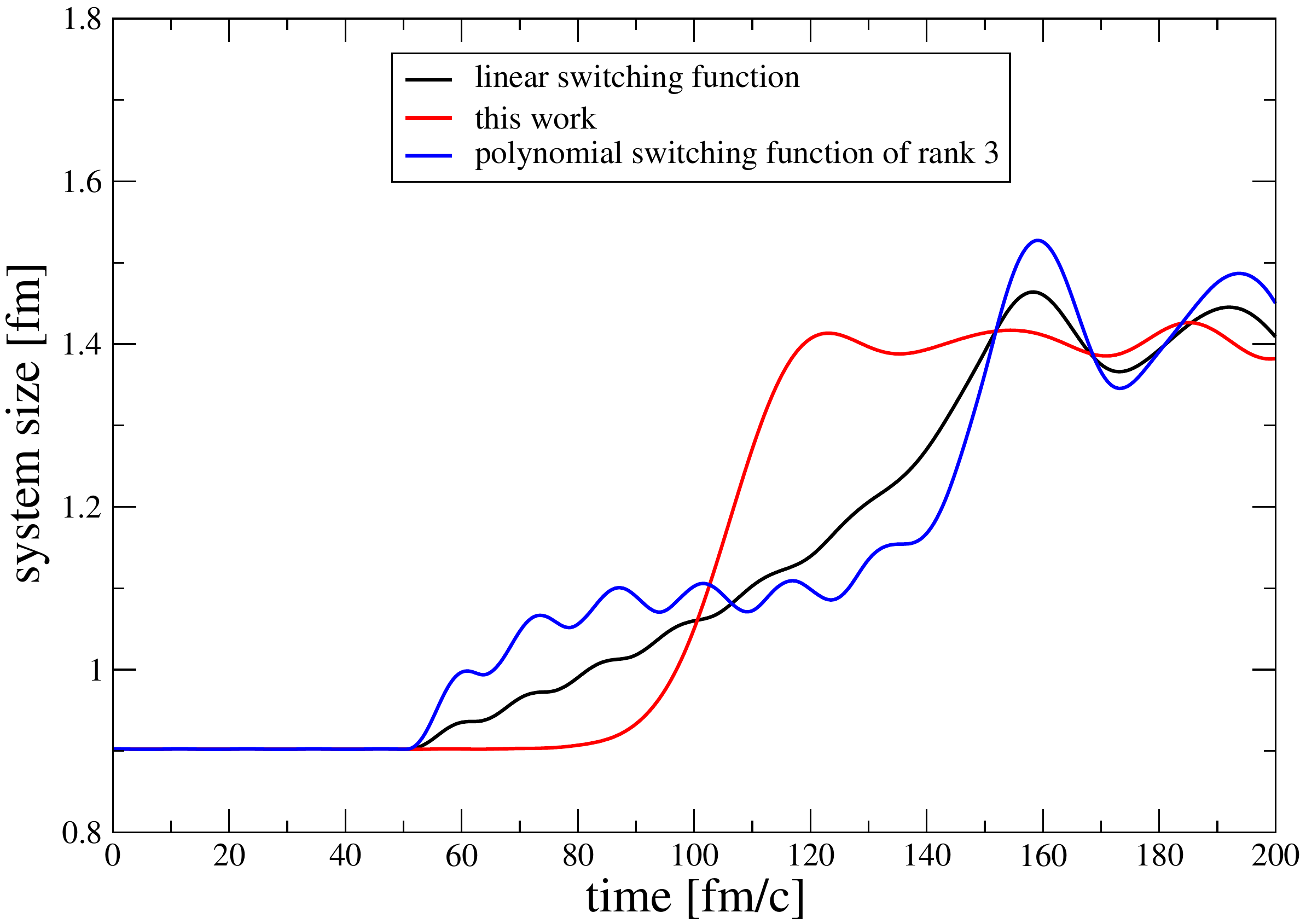}}
  \caption{Time evolution of the size of a system $\langle |x - \langle x \rangle | \rangle $ initialized with one filled HO shell, for different indicated types of switching functions, when the switching is active in the time interval of $[50,150] \, \text{fm}/c$. }
  \label{fig:switchings}
\end{figure}

From~\cite{rios_towards_2011}, our standard switching function in the switching interval from $t_0$ to $t_1$, bracketing $t_s$, is
\begin{equation}\label{eq:F[t]}
F(t)=\frac{f(t-t_s)-f(t_1-t_s)}{f(t_0-t_s)-f(t_1-t_s)} \, ,
\end{equation}
where
\begin{equation}\label{eq:f}
f(t)=\frac{1}{1+e^{t/\tau}}
\end{equation}
and $\tau$ controls the pace of adiabatic switching.  However in the literature there had been a considerable discussion of the effectiveness of different type of functions in the adiabatic switching~\cite{Watanabe}.  For switching to be effective, you want to be able to carry it out over as short time as possible, arriving at a system with the lowest possible energy, exhibiting this, in particular, by being stationary and not spewing particles to the outside.  In Fig.~\ref{fig:switchings}, we show the changes in system size when switching from a system without self-interactions, trapped in a HO potential, to an interacting system within the Green's function approach and using different switching functions, our standard in Eqs.~\eqref{eq:F[t]} and \eqref{eq:f} and some suggested in Ref.~\cite{Watanabe}.  It is apparent that our switching function actually works the best from the choices.  Use of the switching function can be accompanied by the use a cooling friction~\cite{bulgac} that we generalize~\cite{mahzoon17}.

\begin{figure}
  \centerline{\includegraphics[width=.65\textwidth]{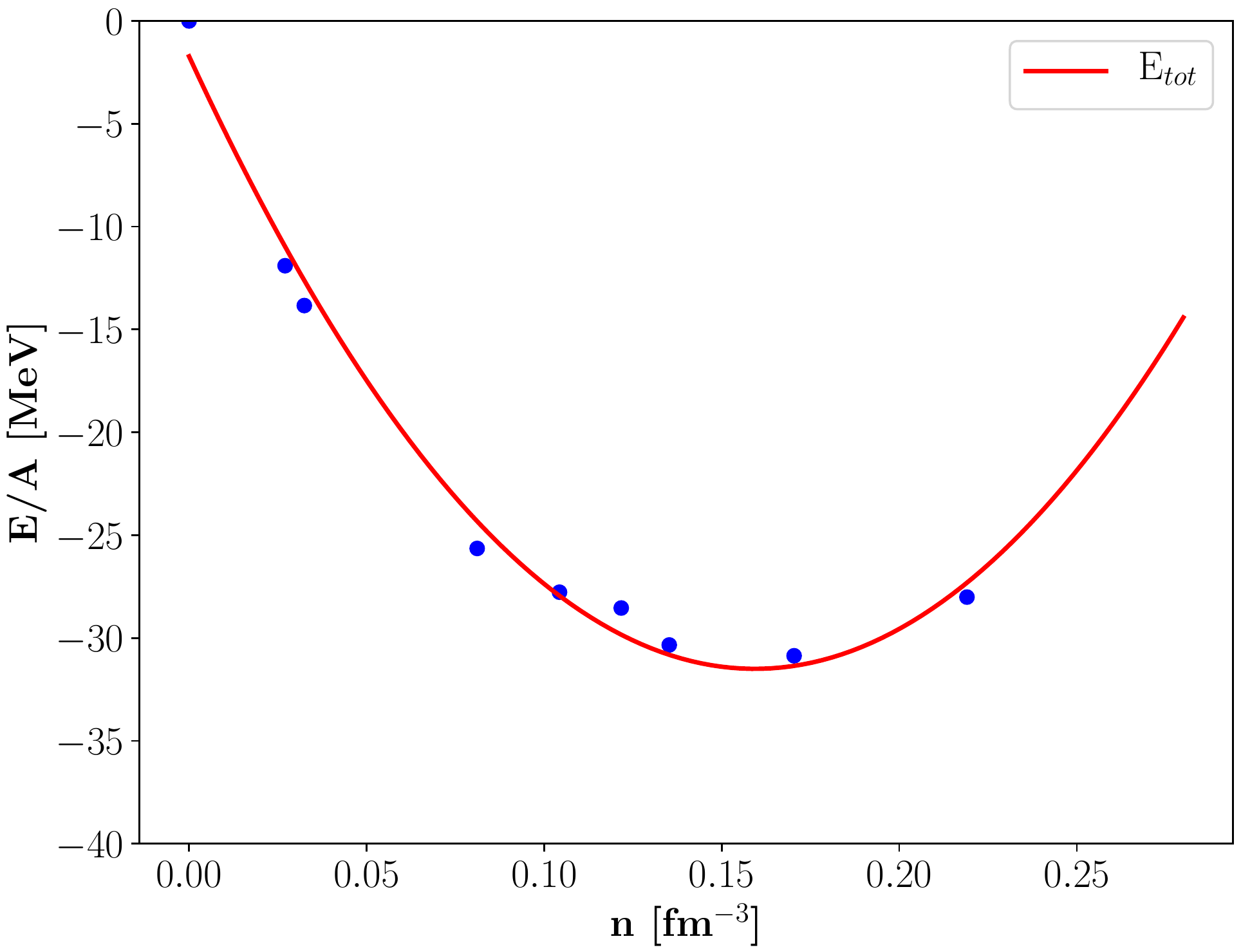}}
  \caption{Net energy per nucleon as a function of 3D density in cold correlated nuclear matter.  Symbols represent values arrived at sample densities and the line represents a smooth interpolation of the results.}
  \label{fig:e_tot_infinite}
\end{figure}

In Fig.~\ref{fig:e_tot_infinite}  we illustrate the energy per nucleon in uniform matter as a function of 3D density for the matter.  We interpret the 1D density in terms of 3D density by following the Hugenholtz-van Hove theorem in Ref.~\cite{rios_towards_2011} and arriving at
\begin{equation}
n_{3D}=\xi \;  n_{1D} \, , \qquad \xi = \sqrt{\frac53}\left (   \frac{\pi n_0^2}{6 \nu^2} \right)^{1/3} \, ,
\end{equation}
where $n_{1D}(x,t)= -i \nu G^< (x \, t;x \,t)$, $\nu=4$ is spin-isospin degeneracy and $n_0 = 0.16 \, \text{fm}^{-3}$ is the normal density.  We aim at a minimum in the energy at the normal density $n_0$.  Numerically there are fluctuations in the energy that we arrive at within the process of switching, such as due to the switching process interplaying with the system oscillations.  A smooth fit to the results helps the eye to identify the location of the minimum.  The relatively low value of the energy per nucleon at the minimum is due to presumed kinetic energy associated with the frozen transverse degrees of freedom missing in the count.

\begin{figure}
  \centerline{\includegraphics[width=.70\textwidth]{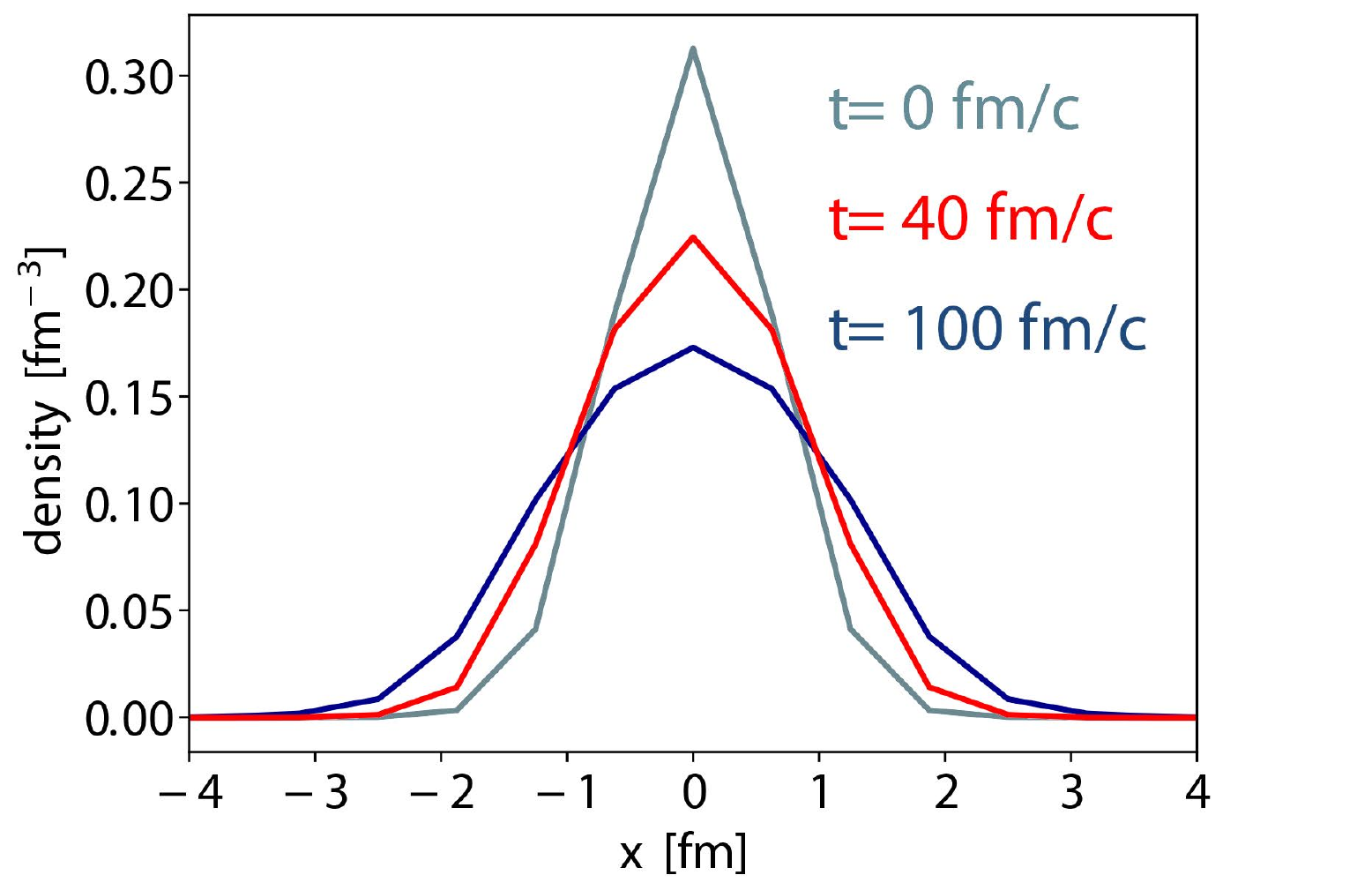}}
  \caption{Time evolution of the density in position space for a system initialized in the first HO shell, when self-interactions are adiabatically switched on and the external potential is relaxed, cf.~Fig.~\ref{fig:switchings}.  Now the switching becomes active from $t=0$ on.}
  \label{fig:density}
\end{figure}

\begin{figure}
  \centerline{\includegraphics[width=.70\textwidth]{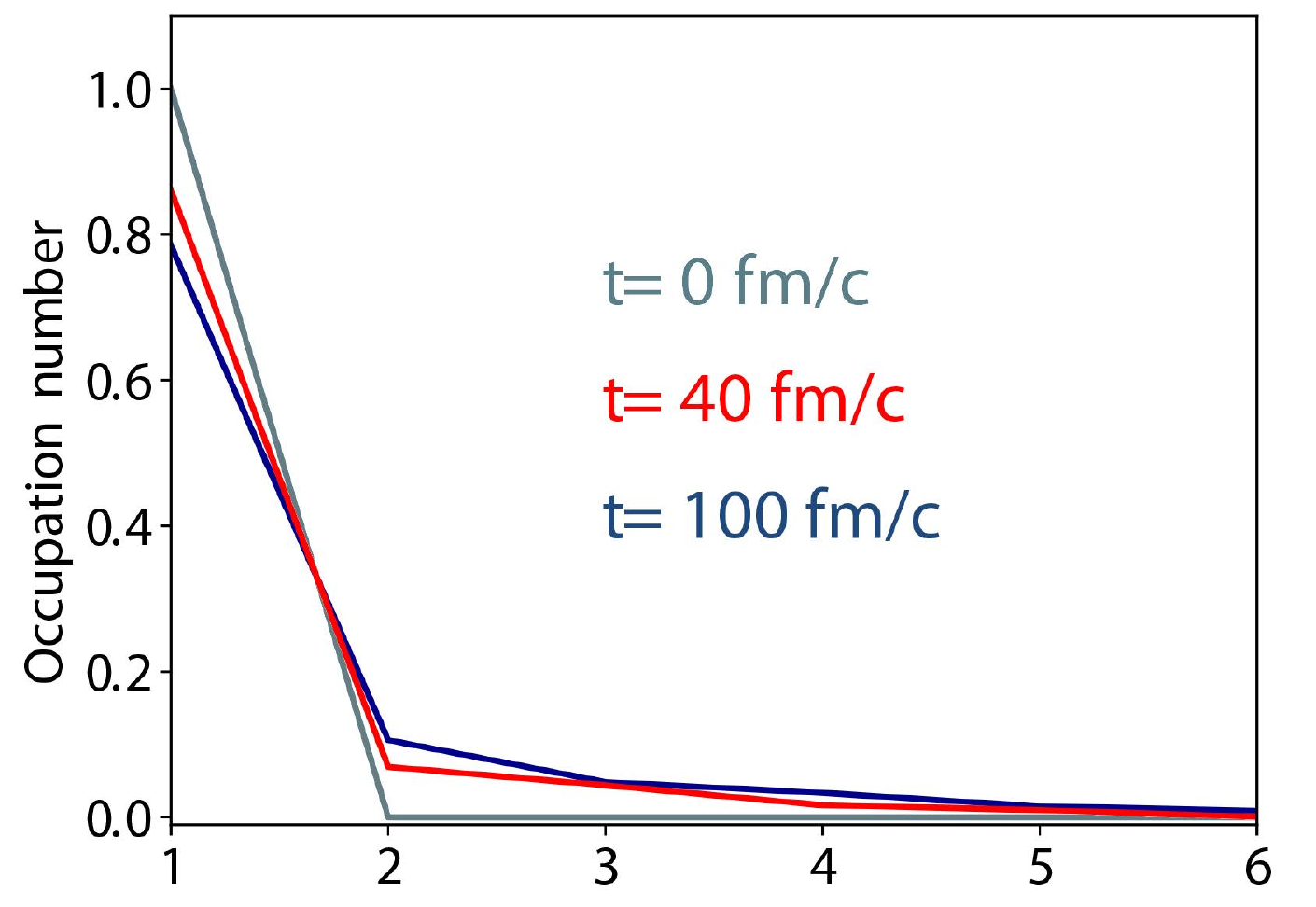}}
  \caption{Occupation of single-particle states from diagonalization of the single-particle density matrix at different times, for a system initialized in the first HO shell, when self-interactions are adiabatically switched on and the external potential is relaxed.  The abscissa is the order of the states by occupation and the lines guide the eye.}
  \label{fig:occupation}
\end{figure}

For energy minimizing at normal density, the finite self-interacting slabs adjust their density profiles so that densities close to the normal are reached at their center.  In Fig.~\ref{fig:density} we show the evolution of the density for a slab initialized in the first HO shell, when interactions are switched on.  The changes in the size for that slab were already shown in Fig.~\ref{fig:switchings}.  Finally, in Fig.~\ref{fig:occupation} we show how the occupation of single-particle states changes during the transformation of interactions.  For an uncorrelated state, in the case of external potential or mean-field only, the occupation in the lowest-energy state is either 1 or 0.  In uniform matter, the states are occupied up to Fermi momentum when there are no correlations and pronounced tails in momentum develop when correlations are switched on.

\section{Conclusions}

We have carried out early steps in advancing the dynamics of correlated nuclear systems following the nonequilibrium Green's function method.  We prepared correlated infinite and finite nuclear systems in 1D through adiabatic switching.  We explored the use of different switching functions, initial conditions and cooling friction, arriving at reasonably stationary and cold states for initiating the dynamics of interest.  In parallel, we developed a suitable combination of mean-field and residual interactions for describing the initial states and the dynamics.

\section{Acknowledgments}

The authors benefited from discussions with Brent Barker and Hao Lin.  This work was supported by the U.S.\ National Science Foundation under Grant PHY-1520971.

%

%
%

\bibliography{copr17}

\end{document}